# Tactical Edge IoT in Defense and National Security

**Paula Fraga-Lamas[1]\*** and **Tiago M. Fernández-Caramés[2]**

[1]*CITIC Research Center, Department of Computer Engineering, Universidade da Coruña, 15071, A Coruña, Spain*
[2]*CITIC Research Center, Department of Computer Engineering, Universidade da Coruña, 15071, A Coruña, Spain*

*Corresponding Author: Paula Fraga-Lamas; paula.fraga@udc.es

**Abstract:** The deployment of Internet of Things (IoT) systems in Defense and National Security faces some limitations that can be addressed with Edge Computing approaches. The Edge Computing and IoT paradigms combined bring potential benefits, since they confront the limitations of traditional centralized cloud computing approaches, which enable easy scalability, real-time applications or mobility support, but whose use poses certain risks in aspects like cybersecurity. This chapter identifies scenarios in which Defense and National Security can leverage Commercial Off-The-Shelf (COTS) Edge IoT capabilities to deliver greater survivability to warfighters or first responders, while lowering costs and increasing operational efficiency and effectiveness. In addition, it presents the general design of a Tactical Edge IoT communications architecture, it identifies the open challenges for a widespread adoption and provides research guidelines and some recommendations for enabling cost-effective Edge IoT for Defense and National

Security.

# 1.1. Introduction

The Internet of Things (IoT) is a distributed system that generates value through data by allowing heterogeneous physical objects to share information and coordinate decision-making. In industrial applications, IoT leads to significant improvements in efficiency and transparency in product development and distribution, as well as in supply chain tracking. It also influences how critical infrastructures are managed and maintained in a wide range of sectors [1], including manufacturing, transportation, logistics, human health and productivity, energy and utilities, home and environment management, or autonomous vehicles. Moreover, IoT redefines how human-machine interactions are performed, being capable of improving equipment performance and enhancing workforce safety [2].

According to a report of McKinsey (November 2021) [3], the potential economic value of IoT will grow from \$5.5 trillion to \$12.6 trillion by 2030. With respect to Machine-to-Machine (M2M) communications, the market is expected to reach \$83.23 billion by 2030, increasing at a compound annual growth rate of 23.2 % [4]. Ericsson states that in 2027 [5], broadband IoT connections with 4G will account for 40 % of cellular IoT connections (related to use cases with high throughput, low latency and large data requirements), while massive IoT connections (associated with use cases that involve a large number of low-complexity, low-throughput and low-cost devices with extended battery life) are expected to account for 51 % of all cellular IoT connections.

Tactical scenarios are mainly characterized by limited resources, high levels of stress and a lack of stability (e.g., they can be highly dynamic, complex and hostile, or mission-critical environments). Military and Public Safety (PS) agents are increasingly relying on IoT to support their tactical missions. Such IoT systems are lightweight and sufficiently powerful to run a range of applications to assist dismounted soldiers or vehicles on the move. Nevertheless, they are connected through wireless tactical networks with restricted bandwidth, reachability, reliability, and latency. As a result, tactical network nodes cannot just rely on secure access to cloud computing services to work. Instead, they must look into other options for leveraging in-situ resources. The concept of Edge IoT involves the use of IoT devices with Edge Computing capabilities, where computation takes place at the network edge. Edge Computing has the potential to improve critical aspects of tactical environments: survivability, resilience, network connectivity, trust and ease of deployment. Such capabilities can lead to an enhanced situational awareness and decision-making at the edge. Therefore, Tactical Edge IoT can help the military and PS agents in adapting to certain environments where adversaries are located in increasingly complex tactical scenarios.

This chapter continues the work presented five years ago in [6], where the authors reviewed the future role of IoT for Defense and PS. Thus, this chapter departs from such background knowledge, updating it to provide a comprehensive approach to Edge IoT applied to Defense and PS.

The remainder of this chapter is organized as follows. Section 1.2 introduces the essential concepts that will be used in the chapter. Section 1.3 reviews the opportunities created by current Commercial Off-The-Shelf (COTS) Edge IoT applications for tactical environments. Section 1.4 presents some promising scenarios for Tactical Edge IoT. Section 1.5 overviews the general design of a Tactical Edge IoT communications architecture. Section 1.6 outlines challenges that hinder the adoption of Tactical Edge IoT technologies and introduces some recommendations for further research. Finally, Section 1.7 is dedicated to conclusions.

# 1.2. Background

## 1.2.1. Tactical Edge IoT drivers

IoT represents the convergence of different disciplines (e.g., electronics, sensors, networks, computing, communications, signal processing or artificial intelligence). As a consequence, five main drivers related with these disciplines are fostering IoT rapid expansion. The first driver is the ever-increasing miniaturization and lower cost of powerful microelectronics (e.g., transducers, receivers or processing units (e.g., micro-controllers, micro-processors, System-on-a-Chip (SoCs), Field-Programmable Gate Arrays (FPGAs), Graphics Processing Units (GPUs), Application-Specific Integrated Circuits (ASICs)).

The second driver is the rapid growth and development of wireless communications systems, from auto-identification and traceability technologies [7, 8] to broadband solutions [9], which are currently being linked to the expansion of 5G/6G wireless connectivity [10, 11, 12, 13].

The third driver is the increase on available data storage and computational capabilities offered by SoC hardware, which keeps on improving at a rapid rate.

The fourth factor is the guarantee of a compute continuum from IoT to the edge and to the cloud [14], which enables the creation of hyper-distributed intelligent IoT applications. Most IoT applications are currently implemented on cloud computing based platforms that enable centralized processing and data storage. However, the cloud itself can be seen as a point of failure, since it can be disrupted by attacks or maintenance tasks, which prevent the access to the deployed centralized services and thus block the whole system. In addition, if an IoT system consists of a high number of connected devices, they will probably generate a lot of data exchanges with the cloud, which will derive into the saturation of the cloud if it is not scaled properly. Additionally, these cloud-based solutions have an inherent high latency and energy consumption.

Due to the aforementioned constraints, novel computing paradigms have emerged thanks to the improvement of IoT end-nodes, which are getting more powerful and efficient, and they have the computational capabilities required to implement end-to-end security mechanisms and high-security cryptographic cipher suites [15]. The main goal of some of the most relevant new paradigms is to process most of the data as close as possible to where end devices are located. Thus, the devices of the higher layers are freed from part of their data processing tasks and the amount of data exchanged with the end devices is reduced significantly [16]. For instance, Mist Computing is a paradigm in which data processing capabilities are moved from gateways to end devices. Edge Computing [17], offloads the cloud from tasks that can be handled by devices at the network edge, close to the end IoT nodes. Fog Computing [18] uses low-power devices on the edge for data acquisition, but data processing is performed at gateway devices. Cloudlets [19] make use of high-end computers that perform heavy processing tasks on the edge [20]. A review of the state-of-the-art on the pervasive edge computing paradigm and its applications to Industrial IoT (IIoT) can be found in [21].

The bandwidth, latency, security and decentralization requirements of today's modern tactical applications are incompatible with centralized clouds directly connected to a large number of IoT devices. As a result, the centralized cloud computing landscape is rapidly becoming distributed and heterogeneous. The compute continuum is created by combining centralized cloud-based coordination and control with edge devices placed near IoT sensors and actuators. In addition, edge devices enable computing to be moved closer to the point where data are generated, thus reducing latency, increasing overall throughput, and improving security.

Finally, the fifth factor is the availability and rapid evolution of enabling technologies like Augmented/Mixed Reality (AR/MR) [22], Virtual Reality (VR) [23], quantum computing [24], Cyber-Physical Systems and digital twins [25], blockchain [26] and Distributed Ledger Technologies (DLTs), Unmanned Aerial Vehicles (UAVs) [27, 28], or Artificial Intelligence (AI) [29]. Specifically, the latter is related to the upsurge of innovative software solutions that involve large data processing in areas like AI/Machine Learning (ML) [30], Big Data or analytics. Particularly, recent AI-enabled applications rely on supervised learning (where models are previously trained and then used), unsupervised learning (where data are fed into a system to extract data patterns), reinforcement learning (where real-time data are used to adjust the model) or federated learning (in which locally trained models are gathered to generate a more complete one while preserving data privacy).

IoT devices usually include transducers to collect large amounts of data on physical parameters. Such information is then processed by using some of the aforementioned technologies, which are mainly focused on data analytics and on extracting data insights. The knowledge gained from such an analysis can be applied for monitoring, automation, control, and prediction.

There is a lot of research in distributed and reliable ML models that take advantage of edge resources. For example, Park et al. [31] describe the transition from cloud-based training and inference towards Edge ML. Zhou et al. [32] combine edge and cloud approaches to guarantee the compute continuum. In [33] the authors investigate the convergence of IoT and AI from the perspective of a collaboration between edge and cloud. Deng et al. [34] provide insights into two main research directions: AI for the edge (also known as Intelligence-enabled Edge Computing) (which focuses on the use of AI for constrained optimization problems in Edge Computing) and AI on the edge (which studies distributed approaches that run AI models efficiently on the edge). Thus, new types of design trade-offs have to be reached, including factors such as energy efficiency, latency, privacy or security.

The previously mentioned five drivers, which can be found in the different layers of the IoT technology stack, will lead the next-generation of IoT networks, notably Edge Computing-based IoT networks, or Edge IoT, as it will be called from now on in this chapter.

## 1.2.2. Defense and Public Safety

Defense and PS entities are essential for preserving security and when it is necessary to react to emergency situations and natural disasters. Previous literature has also used the term Public Protection Disaster Relief (PPDR) instead of PS. Such literature mentions PPDR radio communications as a combination of two important areas of emergency response [6]:

- Public Protection (PP): this kind of communications are used by entities dedicated to maintain law and order, to protect life and property, and to deal with emergency situations.

- Disaster relief (DR): these types of communications are used by organizations that fight societal problems that threat human life. Such problems are usually related to health, economic difficulties, or environmental issues, which can be derived from accidents, sudden natural events, or long-term human activities.

PS agents include police officers, firefighters, border and custom guards, coast guards, medical responders, transportation agents and other organizations that are among the first ones to go to scenarios where critical situations occur. The relationships between such organizations depend on the specific situation, context and legislation. Figure 1.1 gives an outline of their primary function and the relationships among them [6].

Crisis management aims to minimize the impact and damage to people and property, ensuring communication capabilities in difficult circumstances where critical infrastructure is frequently degraded or destroyed. It requires a set of capabilities described by standardization bodies (e.g., TCCA [35] or ETSI [36]). For instance, SAFECOM Guidance for Fiscal Year 2021 [37] includes a compilation of emergency communications systems and technical standards.

Natural disasters and other emergencies (e.g., COVID-19 pandemic) are frequently unforeseen events that cause panic among civilians and disrupt existing resources. In such cases, various PS organizations may be involved in order to respond to large-scale disasters, when civil communications infrastructures

| PS agents | Main responsibilities | Main scenarios |
|---|---|---|
| 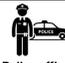 Police officers | Law enforcement and citizen protection: maintain order and secure unpredictable areas, prevent and investigate crime. | Urban/rural environments, major events and border areas. |
| 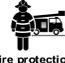 Fire protection | Law enforcement, environmental protection, search and rescue, fire-fighting and fire safety, humanitarian services, management of hazardous materials, mass decontamination. | Urban/rural environments, ports and airports. |
| 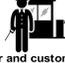 Border and custom guards | Land border control against criminal activity, monitoring of entering goods, and control of illegal immigration. They may be involved in cross-national DR. | Rural environments and border security areas (green border). |
| 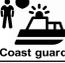 Coast guards | Law enforcement, environmental protection, search and rescue, protection of coastal waters, criminal interdiction, disaster relief, humanitarian assistance. They may be involved in cross-national DR (e.g., earthquake, flooding). | Border areas (Blue border) and ports. |
| 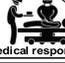 Medical responders | Emergency medical responders to provide critical and life-saving care of sick and injured citizens, as well as to transfer people to a safe environment. | A broad range of scenarios. |
| 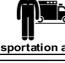 Transportation agents | Law enforcement responsible for the protection of critical transportation infrastructure like roads, railways or airports. | Urban/rural environments. |
| 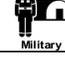 Military | National defense policy and protection of national security. They may support other PS organizations in case of major critical events (e.g., terrorist acts, emergency medical services). | A broad range of scenarios. |

**Figure 1.1:** Description of the main PS agents.

must be operational to send information and warnings to citizens.

In addition, when coordinating DR efforts and establishing situational awareness (SA), PS agents need to exchange data in a timely manner.

Furthermore, interoperability issues may be problem for fulfilling certain security requirements, like the ones related to communications and data protection. To create and maintain a Common Operational Picture (COP) among PS agencies, as well as between field and central Command and Control (C2), different types of data coming from heterogeneous devices must be shared.

The remainder of this chapter will be mainly focused on Tactical IoT with Edge Computing capabilities from a military standpoint. This is due to the fact that such a perspective allows for covering some of the most critical situations and challenging scenarios while fulfilling technical and operational requirements of PS organizations.

In the last years, extensive research has been published focusing on evolving PS communications [38, 39, 40, 41, 42, 43, 44]. Some papers focused on the utilization of LTE and 5G/6G for implementing advanced PS networks [45, 46, 47, 48, 49] or specifically, in device-to-device (D2D) communications [50, 51, 52, 53]. In addition, extensive research has been published in the optimal deployment of Unmanned Aircraft Systems (UAS) for PS networks [54, 55, 56, 57, 58, 59].

There are multiple articles that discuss the diverse parameters that impact the IoT technologies used in Defense and PS applications. For instance, in [60] the authors describe a fault detection mechanism that makes use of a military network divided into clusters. Other authors have proposed layered architectures and discussed different applications related to, for instance, weapon control solutions [61]. A different research line is presented in [62], where the authors present a multi-level authentication service that is lightweight, centralized on a cloud, scable and that considers the timing constraints of the IoT devices used by PS responders.

Nevertheless, there are not many articles that deal with the merge of Edge Computing approaches and IoT devices in tactical environments to comply with the stringent and critical Quality of Service (QoS) requirements of the battlefield or C2. For example, Singh et al. [63] present an IoT health platform for the military. Such a platform includes a semantic-edge network model that serves both tactical and non-tactical information. A more sophisticated approach is described in [64], where the authors propose an IoT framework to monitor troops. The framework is able to enhance decisions that need to be taken during campaigns or battles. The framework uses Mist Computing and ML to further minimize delays. Such a framework is evaluated through simulations of EdgeCloudSim, which show average low network latencies (0.01 s) and failure rates (0.25 %) with a high QoS. The main shortcoming of the system includes its inability to perform successfully with an exponential increase in the number of mobile devices. To overcome to a certain extent this constraint, clusters of mobile devices were deployed by the researchers in a specific area and then applied K-Nearest Neighbours (KNN) to the cluster small number of mobile devices. In addition, the system uses distributed data storage and decentralized data analytics. Another relevant publication is a PhD research [65] that demonstrates how federated learning can be adapted to fit in Edge Computing devices that are connected to a tactical data link.

## 1.3. Compelling COTS Edge IoT Applications

PS operations are currently carried out in complicated, dynamic and often unpredictable circumstances. In particular, military and warfare scenarios have evolved substantially in recent years. As a result, there is a pressing need for military technology to evolve at a fast pace [66]. While military contractors and manufacturers develop new and improved technologies, civilian technology advances at a much faster pace. Moreover, the associated expenses (for development and acquisition) in Defense are substantially greater, and the cycle time is much longer than COTS technology.

This section overviews some COTS Edge IoT applications that may be relevant for Defense and PS environments. Two kinds of deployments must be considered: the ones aimed at creating ad-hoc IoT for the military field and the ones related to already-developed civil deployments that need to be protected (e.g., smart cities).

- Transportation: the authors of [67] propose a blockchain-enabled Edge IoT framework for maritime transportation systems. Blockchain and smart contracts help in the validation of each block's transactions at edge nodes by estimating their lifetime and trustworthiness, as well as mitigating many forms of security threats. Support Vector Machine (SVM) and Convolutional Neural Network (CNN) are used to predict the number of malicious entities and increase the prediction accuracy of vessel monitoring units.

- Energy efficiency: In [68] the authors study how to leverage heterogeneous computation resources at the Edge to optimize energy efficiency while satisfying the delay requirements of Mobile Edge Computing (MEC) scenarios. As a result, they come up with an iterative framework that considers two sub-problems: transmission power allocation and computation offloading. The authors indicate as a limitation that their work does not consider heterogeneous edge servers deployed with several types of computation resources (e.g., CPU, GPU). An energy-efficient MEC is also presented in [69], specifically for the case of UAVs and their specific constraints (e.g., limited flight time, power constraints).

- Supply chain/logistics: In [70] it is presented a lightweight authentication protocol for supply chains in a 5G MEC scenario that makes use of blockchain and Radio Frequency Identification (RFID).

- Smart cities: Khan et al. [71] review the state-of-the-art of Edge Computing applications for smart cities. The authors devise a comprehensive taxonomy and discuss the main requirements. Open challenges are outlined to guide further research.

The above-mentioned examples illustrate how the civil sector is leveraging Edge IoT to enable new business models. Concerns, like cybersecurity, cooperative load balancing, collaborative Edge Computing [71], or green IoT [72], still remain as open research.

# 1.4. Target Scenarios for Tactical Edge IoT

The Network-Centric Warfare (NCW) paradigm [73] connects battlefield assets back to headquarters. Such a concept provides benefits by facilitating the exchange of information among users in a secure and timely way. In addition, the NCW paradigm combines three domains: the physical domain, which generates data where events and actions take place; the information domain, which transmits and stores data; and the cognitive domain, which processes and analyzes data to enable decision-making and mission planning. NCW three domains correspond to the underpinnings of today's commercial Edge IoT.

In network-centric C2 operations, responsibility is delegated to the battlefield edge [74], creating the so-called Internet of Battlefield Things (IoBT), which can bring together everything on the battlefield that can aid in making informed decisions. However, these dynamics require network paradigms that can ensure network efficiency. In [74] the authors bring together Information-Centric Networking (ICN) with Software-Defined Networking (SDN) to fulfill such requirements.

This section analyzes some of the most relevant Tactical IoT scenarios with Edge services, which are summarized in Figure 1.2. Applications for Command, Control, Communications, Computers, Intelligence, Surveillance and Reconnaissance (C4ISR) and fire-control systems have dominated the adoption of IoT-related technology for Defense and PS, since sensors are primarily used to collect and communicate data in order to improve C2. Although IoT and Edge Computing technologies have previously been used for applications related to logistics and training, their integration with other systems is often limited.

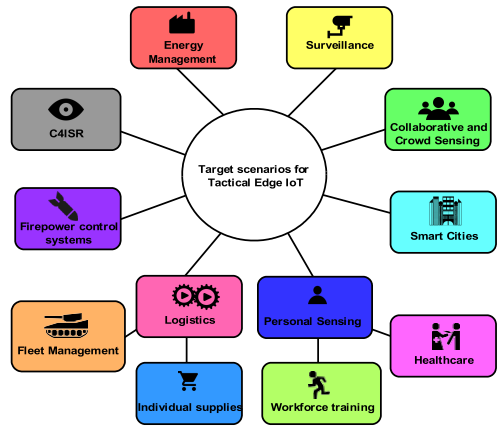

**Figure 1.2:** Target scenarios for Tactical Edge IoT in defense and public safety.

As it was previously stated, Edge IoT capabilities can be used to provide superior SA in the battlefield. Commanders can make decisions based on real-time analysis derived from the integration of AI/ML extracted data from unmanned/manned sensors and field reports. Ground-based sensors and cameras, as well as human or unmanned devices, vehicles or soldiers, provide a wide range of information to commanders. The mentioned IoT devices are able to scan the mission environment and then send information to an Edge Computing server, which can be in a forward base. Part of such information may be collected by a Command Center, where it will be processed and fused with information from other sources.

### 1.4.1. C4ISR

C4ISR systems provide advanced situational awareness by deploying millions of sensors across a variety of platforms. Surveillance satellites, aerial platforms, UAVs, ground stations, and soldiers in the battlefield collect different types of data (e.g., radar, infrared, video). Such information is collected, processed and stored in a platform, which manages the data up and down the chain of command. These platforms provide a COP, allowing for better battlefield coordination and control.

Central operation centers, which receive data from platforms, provide comprehensive situational awareness to high-level military echelons. Lower levels (e.g., platoons and soldiers) have access to data in their specific areas. Combat pilots, for example, receive prioritized data streams that are combined with data from their own sensor systems.

Jiao et al. [75] present a mechanism for moving services from cloud-based systems to combat platforms, thus allowing an effective deployment of C4ISR systems in scenarios where communications bandwidth is limited. Moreover, the researchers propose a two-level search algorithm based on improved quantum evolution, which also combines path planning and pair-wise exchange techniques.

## 1.4.2. Firepower Control Systems

Sensor networks and advanced AI/ML analytics in fire-control systems enable fully automated reactions to real-time threats and pinpoint accuracy delivery of firepower. Smart weapons can also follow moving targets or be diverted in mid-flight.

### 1.4.3.  Logistics

Multiple low-level sensors are employed in logistics in the Defense sector. Their deployment is mainly focused on safe contexts (e.g., non-combat settings) with infrastructure and human interaction to improve back-end processes. The examples in the following subsections are divided into two main categories: fleet management and individual supplies.

### 1.4.3.1.  Fleet Management

On-board sensors within aircraft and ground vehicle fleets can be used, for instance, to assess the system performance, to monitor part status, or to evaluate the condition of a vehicle and its subsystems. In addition, it is possible to alert users when certain goods (such as fuel or oil) need to be replenished or when a failure is expected. With an Edge Computing approach, sensors would send out real-time alerts, potentially lowering the danger and the probability of a catastrophic event. Although Edge IoT deployment has upfront expenses, it can result in considerable long-term savings across the different processes.

Defense has an opportunity to leverage the advances in data sharing of the automotive industry (e.g., industrial IoT-connected vehicles or autonomous vehicles).

There are other relevant parameters that can be useful in real-time fleet management, like the location of the vehicles, their speed, engine status, fuel efficiency, total weight or the carried load. In addition, in cases when shipments need to be monitored, the location and status of the containers can be tracked in order to detect potential problems.

Regarding aircrafts, current jet engines include sensors that collected huge amounts of data per flight (in the order of terabytes) [76]. The real-time edge analytics information, when paired with in-flight data, can help reduce fuel expenditures, detect small defects, and minimize journey times. It also allows for providing preventive maintenance, which results into longer life cycles (since failures are reduced or totally prevented) and into dedicating less time to repairs.

### 1.4.3.2. Individual Supplies

Individual supplies can be tracked using standardized barcodes, RFID tags, specific smart labels [77] or a wide range of auto-id and traceability technologies [7]. Edge IoT allows the military to monitor the status of the supply chain in real time, so it enables knowing when goods are delivered, moved, deployed, or consumed.

In addition, real-time tracking can be useful for soldiers, when it is necessary to take a proactive approach to logistics. Critical soldier supplies such as water, food, batteries, or bullets can be tracked and alarms provided if a resupply is required or when an unexpected event happens. Information aggregated from military units (e.g., groups of soldiers, companies or battalions) can be examined with AI/ML techniques on the edge for further real-time supply improvements for tactical and emergency forces by taking into account context awareness (e.g., characteristics such as the surroundings, body type, daily intake or weather information).

## 1.4.4. Smart City Operations

Existing IoT smart city infrastructure can be potentially employed for military operations [78]. For instance, hazardous chemicals can be monitored through environmental sensors, while different sensors have the capacity to track the behaviour of people that act suspiciously.

Taking advantage of real-time data offered by current infrastructure could be crucial from some operations. Nevertheless, security risks, such as equipment sabotage or false information, may arise. According to [79], attacks may fall into main four categories: (1) firewalls, software patches, and system design; (2) malware, security policies, and human factors; (3) third-party chains and insider threats; and (4) database schemas and encryption technologies. For instance, an investigation of the different security threats of multi-access MEC from a physical layer perspective is presented in [80]. Other articles like [81] have previously analyzed privacy security in Edge Computing and then proposed an efficient intelligent offloading method for smart cities that preserves privacy.

## 1.4.5. Soldier Healthcare and Workforce Training

Wearable technology embedded within military equipment (e.g., weapon systems, combat suits) allows for the ubiquitous tracking of physical activity and the collection of operational context data as well as biometrics [82]. Inferring and monitoring physical or psychological conditions from context-aware information in real-time and taking preventive actions could be critical. For example, soldiers can be notified of different events (e.g., dehydration, sleep deprivation, elevated heart rate, low blood sugar, speech patterns, likelihood of internal injury based on earlier traumas) and, when needed, notifications can be sent to a base hospital so that the medical team take the necessary actions [63].

In addition, two types of sensing can be distinguished: participatory and opportunistic. The latter could be very useful for undercover troops conducting reconnaissance missions in urban areas.

It should be also considered the existence of dismounted soldiers, mainly equipped with a mobile device with an application (e.g., ATAK android app) to receive and send SA data from the tactical operation center [78].

IoT can also be utilized for achieving an improved experience for training, gaming, and simulation exercises. For example, wearable devices can be used to simulate real-life fighting and to track the workforce [83].

More details on IoT equipment for soldiers can be found in [6].

## 1.4.6. Collaborative and Crowd Sensing

Collaborative sensing is a way of sharing the information collected from sensors across mobile devices, often by making use of reliable short-range communications. To supplement their own sensing methods, IoT nodes could use additional sensors. The data fusion information can be made available to soldiers once potential security threats (such as trust and authentication) have been handled.

By pairing sensors with mission assignments, IoT can make ad-hoc ISR missions easier. As a result, sensors and platforms would not need to be overly equipped to handle missions since they can rely on collaborative Edge Computing-based sensing capabilities to accommodate specific needs on demand.

To create a COP, resource-rich devices may collect data from multiple sources. Much of these data might be stored and processed locally. As a result, Edge Computing functions at a higher level would help in reducing response times and the need for backhaul connections.

Crowdsensing has the potential to be a low-cost tool for flexible real-time monitoring of broad regions, complementing services that may be available in smart cities. Nevertheless, security has to be carefully considered [84]. Data validation is another issue that is further hampered by the great heterogeneity of the devices. Device capabilities and performance may change over time. Low battery levels on a smartphone could result in occasional GPS position updates, resulting in geo-tagging inaccuracies. Oversampling and filtering outlier values are common approaches to ensuring data quality. In addition, reputation methods that give trustworthy sensing for PS can be used [85].

Moreover, inherent privacy issues may jeopardize crowdsensing services. For instance, when monitoring certain soldier activities, geo-tagging and timestamping may be needed, which can lead to the revelation of the location of such soldiers. The metadata acquired about devices by performing sensing activities is another privacy concern.

Furthermore, when processing data, the existence of attacks against AI/ML systems shall be carefully considered [86].

## 1.4.7. Energy Management

The use of real-time IoT data and predictive algorithms can aid in a better understanding of usage trends and drastically reduce military energy expenses.

## 1.4.8. Smart surveillance

Real-time remote facility monitoring for security threats is enabled by security cameras and sensors, as well as advanced AI/ML image processing and pattern recognition software. In the case of maritime environments, it is possible to embed different types of sensors into helicopters, airplanes, UAVs, or ships. Thus, IoT solutions allow for monitoring marine activities and ship traffic over large areas, as well as sensing environmental conditions or the status of dangerous oil cargos.

Hazardous environmental parameters can also be monitored by Edge IoT systems, which can alert users fast when certain conditions are met.

# 1.5. Communications Architecture

After reviewing the current state of the art and the requirements of potential Tactical Edge IoT applications [6], a communications architecture like the one depicted in Figure 1.3 can be devised. As it is shown in such a Figure, it is a four-layer Edge Computing architecture that supports data collection and that is composed by the following main components:

**Figure 1.3:** Communications architecture of a Tactical Edge IoT system

¬ The node layer is at the bottom. Such an IoT Device Layer includes IoT devices that a dismounted soldier might carry (e.g., RFID tags, wearables, body sensors, environmental sensors, AR/VR devices, and handheld radio), as well as other systems like UAVs, vehicles on the move or sensors deployed on the mission scenario.

ï Each IoT device exchanges data via a wireless connection with a Tactical Edge Layer gateway, which is usually the one that is physically closest to. All local fog gateways are part of a Fog Computing sub-layer (Access Sublayer) that provides services with low latency requirements such as sensor fusion, AI/ML services, position services, or data caching for streaming content to AR/VR devices. Although most of the time only one gateway provides fog services to a single IoT device, the Access Sublayer gateways can cooperate to provide complex and more sophisticated services (e.g., to distribute compute-intensive tasks, to make it easier the collaboration between remote IoT devices). Local gateways can be Single-Board Computers (SBCs) or similar devices, which are reduced-size low-cost computers that can be quickly deployed in a tactical scenario. As a result, the architecture assumes that the devices that act as gateways can be easily scattered throughout the battlefield. In addition, the Tactical Edge Layer contains cloudlets, which can perform compute-intensive activities like rendering or AI/ML processing. The cloudlet response latency is significantly lower than that of standard cloud computing systems because it is close to the IoT devices that request its services.

ï In the upper layer, the top-level gateway is the point of entry to the lower layers, while the other gateways provide different services or share data among them to reduce the latency response from the cloud, acting as backbone gateways.

ï Finally, located at the top of Figure 1.3, is the internal cloud, where the services that demand more processing power are executed. Third-party systems, part of the military IT core, are also connected to the cloud. The cloud makes available certain services to remote users (e.g., commanders that need to access the stored information in remote headquarters).

Although it is not required for the basic operation of a Tactical Edge IoT system, a blockchain or DLT provides additional benefits like trustworthiness, redundancy, or security [87]. Moreover, smart contracts can be implemented on a blockchain, allowing for the automation of certain operations in response to detected events.

# 1.6. Main Challenges and Recommendations

Although deployments of Tactical Edge IoT solutions for real-world scenarios have already begun, they face major challenges [78, 88]. Therefore, relevant issues still need to be tackled, like trustworthiness (e.g., algorithm transparency, traceability, privacy, and data integrity); capacity (e.g., communications bandwidth and coverage); security in edge distributed architectures; heterogeneity; or scalability. Figure 1.4 illustrates different open research lines.

Some recommendations can be provided to future researchers:

ï In order to define metadata across very heterogeneous and different domains, ontologies are required [78].

ï With respect to data management, the communications format should include automatic data descriptors allowing for more efficient data management and storage across domains [78].

ǐ Tactical IoT systems must take into account that the architecture various nodes (e.g., mist nodes, Edge Computing devices, and cloudlets) have varying capacities in terms of communications, computation, storage, and power. The communications format must be supported across platforms, which implies that it should not depend on the used hardware or on underlying software. In addition, developers should consider that IoT devices make use of relatively disruptive networks, which have to tolerate faults and should support requirements related to QoS.

ǐ The use of open standards is highly recommended for military and civilian applications, since it is a way for ensuring long-term interoperability.

ǐ Energy efficiency and low-latency architectures for Tactical Edge IoT systems pose multiple challenges. Therefore, developers should consider aspects like the use of low-power communication technologies (e.g., ZigBee, Wi-Fi Hallow, LoRa, LoRaWAN), the smart management of the radio spectrum or the creation of distributed AI-based solutions able to achieve low inference latencies despite the requirements related to training and learning. In addition, renewable energy sources and energy-harvesting techniques should be considered in order to reduce the dependence on traditional energy source like batteries.

ǐ The protocols used for exchanging messages have to support multi-level security mechanisms, since military information often requires to be secured at various levels [78]. Specifically, security mechanisms are needed to protect Tactical Edge IoT systems from attacks at physical, network and application levels, as well as from encryption attacks and software vulnerabilities. Moreover, AI learning processes should be protected against adversarial attacks.

- The deployment of 5G/6G networks supposes significant reductions on latency and increases in uplink/downlink data rates. Such changes will enable to move part of the processing power to the edge, where more powerful Edge Computing devices will be demanded. In addition, the popularization of new devices with native global connectivity will derive into the need for easing the convergence of Tactial Edge IoT systems with 5G/6G networks and other related technologies (e.g., multi-band radios for low-bandwidth scenarios, Mobile Ad-Hoc Networks (MANETS) or the use of defensive countermeasures).

- The military should explore forming a specialized technology group made up of military personnel to test new technologies and gather real-world input early on the development process. This evaluation has the potential to create innovative new uses for Edge IoT devices. Its two-fold purpose would be to find devices and systems with possible applications, as well as to discover new approaches for completing missions employing COTS.

- The delivery of web-based services can be performed through Platform as a Service (PaaS) solutions without developing or managing infrastructure, which results in systems that provide more flexibility and scalability for adjustments and updates. The adoption of PaaS in military environments imposes additional challenges, requiring the implementation of security processes by private contractors.

- The creation of a comprehensive trusted architectures that can fulfill all the military Tactical Edge IoT requirements.

- Governments and Defense may invest in enabling technologies to improve

Tactical Edge IoT implementation. Further integration of other digital enabling technologies (e.g., quantum computing, digital twins, blockchain and DLTs, functional electronics, UAVs, AR/VR) is needed.

ï Further collaboration with private entities is necessary for updating current Edge IoT systems with the latest technologies. Civil companies are hesitant to collaborate with the military due to cultural gaps, as well as differences when managing intellectual property. For example, private entities may consider as much too challenging the development of certain military Tactical Edge IoT solutions, especially when having to deal with complex and demanding operational requirements. In addition, Tactical Edge IoT implementation will require the compromise of all stakeholders.

# 1.7. Conclusions

This chapter analyzed some of the ways of how the Defense industry can take advantage of the commercial Edge IoT transformation. Essential issues were discussed concerning the development of Edge IoT applications for the military and PS domains. Specifically, potential Tactical Edge IoT application scenarios were described, like C4ISR, fire-control systems, logistics, smart cities, health-care, training, crowd sensing, energy management and smart surveillance. In addition, the design of a generic Tactical Edge IoT communications architecture was presented.

Furthermore, it was emphasized the fact that Defense presents additional challenges to COTS Edge IoT systems, mainly posed by tactical surroundings, as well as the inner complex nature of operations and networks. Governments and the Defense sector might gain a competitive advantage by utilizing existing

COTS technologies and business methods. As a result, some recommendations were provided for enabling cost-effective Tactical Edge IoT.

| Identification | Infrastructure | Signal Processing | Energy efficiency | Interoperability | Standardization | Discovery |
|---|---|---|---|---|---|---|
| Identity management<br><br>Open framework for Edge IoT<br><br>Soft Identities<br><br>Semantics<br><br>DNA identifiers<br><br>Convergence of IP and IDs and addressing scheme<br><br>Multimethods, one ID | Cross domain application deployment<br><br>Integrated IoT, multi-application and multi-provider infrastructures<br><br>Discovery mechanisms | Context aware data processing<br><br>Distributed energy efficient data processing<br><br>Cognitive processing<br><br>Ontologies | Energy harvesting<br><br>Power generation in harsh environments<br><br>Biodegradable batteries<br><br>Nano-power processing unit<br><br>Energy recycling | Reduced cost of interoperability<br><br>Open platform for IoT validation<br><br>Dynamic and adaptable interoperability for technical and semantic areas | M2M<br><br>Cross interoperability with heterogeneous networks<br><br>Information sharing<br><br>Interaction | Automatic identification management<br><br>Semantic discovery of sensors<br><br>Cognitive search engines<br><br>Autonomous search engines<br><br>Discovery services |

| Architecture | Software | Hardware | Network | Communications | Security | Applications |
|---|---|---|---|---|---|---|
| Network of networks architectures<br><br>Adaptive and context based architectures<br><br>Self-* properties<br><br>Cognitive and experimental architectures<br><br>Code in tags<br><br>Cyber-physical systems, digital twins | Distributed intelligence<br><br>M2M collaboration environments<br><br>Edge IoT complex data analysis<br><br>Edge IoT intelligent data visualization<br><br>Hybrid IoT<br><br>Human-centered IoT<br><br>Service reliability<br><br>Fully autonomous IoT devices<br><br>Micro operating systems<br><br>Context aware business event generation<br><br>Interoperable ontologies of business events | Smart bio-chemical sensors<br><br>Nano-technology and new materials<br><br>Collaborative tags<br><br>Self-powering sensors<br><br>Molecular sensors<br><br>Transparent and flexible displays<br><br>Biodegradable antennas<br><br>Nano-power processing units<br><br>Multi-protocol frontends<br><br>Collision free air to air protocol and minimum energy protocols<br><br>Multi-band, multi-mode wireless sensor architectures<br><br>Reconfigurable wireless systems<br><br>Micro readers with multi-standard protocols | Self- * networks<br><br>IPv6-based deployments<br><br>Software defined networks<br><br>Service based network<br><br>Authentication<br><br>Robust security based on a combination of ID metrics | Wide spectrum and spectrum aware protocols<br><br>Ultra-low power system on chip, multi-protocol chips<br><br>Multi-functional reconfigurable chips<br><br>On-chip networks and multi-standard RF architectures<br><br>Seamless networks<br><br>Gateway convergence<br><br>Hybrid network technologies convergence<br><br>5G/6G developments<br><br>Collision-resistant algorithms<br><br>Plug-and-play tags, self-repairing tags | User centric context-aware privacy<br><br>Privacy aware data processing<br><br>Security and privacy profiles and policies<br><br>Context centric security<br><br>Homomorphic Encryption, searchable Encryption<br><br>Self-adaptive security mechanisms and protocols<br><br>Access control and accounting schemes<br><br>General attack detection and recovery<br><br>Decentralized self-configuring methods for trust establishment | AI/ML capabilities<br><br>Cross-domain integration and management<br><br>Context-aware adaptation of operation<br><br>Standardization of APIs<br><br>Human-centered interaction |

**Figure 1.4:** Ongoing Tactical Edge IoT research [6].

# Acknowledgments

This work has been funded by the Xunta de Galicia (by grant ED431C 2020/15, and grant ED431G 2019/01 to support the Centro de Investigación de Galicia "CITIC"), the Agencia Estatal de Investigación of Spain (by grants RED2018-102668-T, PID2020-118857RA-I00 and PID2019-104958RB-C42) and ERDF funds of the EU (FEDER Galicia 2014-2020 & AEI/FEDER Programs, UE).